\documentclass[prl,aps,twocolumn,preprintnumbers,showpacs,superscriptaddress]{revtex4}

\usepackage{amsfonts,amssymb,amsmath}            
\usepackage{graphics,graphicx,epsfig}            
\usepackage{amsthm}                              

\def\openone{\leavevmode\hbox{\small1\normalsize\kern-.33em1}}

\newcommand{\be}{\begin{equation}}
\newcommand{\ee}{\end{equation}}
\newcommand{\bea}{\begin{eqnarray}}
\newcommand{\eea}{\end{eqnarray}}

\begin{document}

\title{Mirror Inversion of Quantum States in Linear Registers}

\author{Claudio \surname{Albanese}}
\email[]{claudio.albanese@imperial.ac.uk}
\affiliation{Department of Mathematics
             Imperial College,
             London, SW7 2AZ, UK}
\affiliation{Department of Mathematics,
             National University of Singapore,
             Singapore 117\,543, Singapore}

\author{Matthias \surname{Christandl}}
\email[]{matthias.christandl@qubit.org}
\affiliation{Centre for Quantum Computation, DAMTP,
             University of Cambridge,
             Cambridge CB3 0WA, UK}

\author{Nilanjana \surname{Datta}}
\email[]{n.datta@statslab.cam.ac.uk}
\affiliation{Statistical Laboratory, DPMMS,
             University of Cambridge,
             Cambridge CB3 0WB, UK }

\author{Artur \surname{Ekert}}
\email[]{artur.ekert@qubit.org}
\affiliation{Centre for Quantum Computation, DAMTP,
             University of Cambridge,
             Cambridge CB3 0WA, UK}
\affiliation{Department of Physics,
             National University of Singapore,
             Singapore 117\,542, Singapore}

\begin{abstract}
Transfer of data in linear quantum registers can be significantly
simplified with pre-engineered but not dynamically controlled
inter-qubit couplings. We show how to implement a mirror inversion
of the state of the register in each excitation subspace with
respect to the centre of the register. Our construction is
especially appealing as it requires no dynamical control over
individual inter-qubit interactions. If, however, individual
control of the interactions is available then the mirror inversion
operation can be performed on any substring of qubits in the
register. In this case a sequence of mirror inversions can
generate any permutation of a quantum state of the involved
qubits.
\end{abstract}


\maketitle

The network (circuit) model of quantum computation is justifiably
the most popular model for investigating both computational power
and possible experimental realizations of quantum computers. One
of its many appealing features is the reduction of quantum
computation to prescribed sequences of elementary operations
(quantum logic gates) performed either on individual qubits or on
pairs of qubits~\cite{Deu89}. However, a tacit assumption that
single- and two-qubit operations are much easier to implement than
multi-qubit operations, is not always valid. In fact, there are
potentially interesting technologies, for example, optical
lattices~\cite{MGWRHB03}, arrays of quantum
dots~\cite{Eke95e,BDEJ95,Kan98a,LD98a}, or
NMR~\cite{CFH96a,GC97a}, in which joint operations on several
qubits are relatively easy whereas addressing individual qubits
poses a substantial experimental challenge. Thus it is important
to investigate quantum computation with limited control over
individual qubits. Here we show that transfer of data in quantum
registers can be significantly simplified with pre-engineered but
not dynamically controlled inter-qubit couplings.

It is known that quantum computation could in principle be
performed by a chain of qubits coupled via the Heisenberg or the
XY interactions~\cite{DBKB00}, and that it suffices to control the
qubits collectively~\cite{Ben02}. Such a chain of qubits
represents a quantum register. Further simplifications to this
model have been recently introduced by Zhou {\it et
al.}~\cite{ZZGF02} and by Benjamin and Bose~\cite{BB03}. Still, a
significant number of elementary operations in the process of
computation is delegated to moving around quantum states of
individual qubits. We show how to simplify these operations by
implementing a mirror inversion of a quantum state with respect to
the centre of the chain. More precisely, given a chain of $N+1$
qubits described by the wavefunction $\Psi(s_0, ... s_N)$, where
$s_n=0,1$ denotes the bit values of the $n$th qubit, we show how
to implement the transformation $R$
\be%
R \; \Psi(s_0, s_1...,s_{N-1}, s_N) = (\pm)\Psi(s_N,
s_{N-1},...,s_1, s_0). \label{eq_defineR}
\ee%
Our construction has the advantage that it can be done without
applying any dynamical control to the qubits, it only exploits the
natural dynamics of the chain governed by a pre-engineered mirror
periodic Hamiltonian $H$ such that $\exp({-i T H})=R$ for some
time $T$.

Apart from obvious applications, such as a perfect quantum wire or
a `data bus' linking the two opposite ends of the chain studies of
periodic and mirror-periodic dynamics of chains of spins with
non-homogenous couplings is an interesting subject on its own,
with potential applications outside quantum computation, e.g. in
the design of frequency standards and in mathematical finance.

Consider $N+1$ interacting qubits, or spin--$1/2$ particles, in a
quantum register. We choose the Hamiltonian of the system to be of
the XY type
\begin{equation}
H = \frac{1}{2}\sum_{\ell = 0}^{N-1} J_{\ell} \big(
\sigma^{x}_{\ell} \cdot \sigma^{x}_{\ell+1} + \sigma^{y}_{\ell}
\cdot \sigma^{y}_{\ell+1} \big) - \frac{1}{2} \sum_{\ell=0}^N
h_{\ell} \big(\sigma^{z}_\ell - 1\big), \label{eq_spinham}
\end{equation}
where $J_\ell$ is the coupling strength between the qubits located
at sites $\ell$ and $\ell+1$, and $h_\ell$ is the `Zeeman' energy
of a qubit at site $\ell$. Please note that here $\ell$ labels the
position of a qubit in the register, whereas the three Pauli
matrices are denoted as $\sigma^{x}$, $\sigma^{y}$, and
$\sigma^{z}$.

Now our task is to find the values $J_\ell$ and $h_\ell$ for which
the Hamiltonian $H$ is mirror periodic. The total $z$-component of
the spin, given by \be \sigma_{tot}^{z}:= \sum_{\ell=0}^N
\sigma_\ell^{z} \label{spintot} \ee is conserved, i.e.,
$[\sigma_{tot}^{z}, H]=0.$ Hence the Hilbert space of the register
decomposes into invariant subspaces, each of which is a distinct
eigenspace of the operator $\sigma_{tot}^{z}$. The eigenspace with
eigenvalue $(2M-N-1)/2$ corresponds to exactly $M$ qubits having
bit value $1$. Let us denote this subspace by ${\cal{S}}_M$.

For convenience of our exposition, we adopt here the standard
fermionization technique~\cite{LiebWu}. We will view the register
as a lattice with $N+1$ sites, some of which are occupied by
indistinguishable and non-interacting, spinless fermions. The bit
values $1$ and $0$ indicate the presence and the absence of the
fermion at a given lattice site and the Pauli exclusion principle
prevents two or more fermions to occupy the same site. The
subspace ${\cal{S}}_M$ corresponds to the $M$--fermion sector, in
which $M$ of the $N+1$ lattice sites are occupied by fermions. The
Jordan-Wigner transformation
\be%
a_\ell=
\big(\prod_{k<\ell}\sigma_k^{z}\big)\frac{\sigma_\ell^{x}+i\sigma_\ell^{y}}{2}
, \, a^\dagger_\ell=
\big(\prod_{k<\ell}\sigma_k^{z}\big)\frac{\sigma_\ell^{x}-i\sigma_\ell^{y}}{2}
\ee%
allows to rewrite the Hamiltonian (\ref{eq_spinham}) in the second
quantization form using the fermionic operators $a_\ell$ and
$a^\dagger_\ell$,
\begin{equation}
H =\sum_{\ell=0}^{N-1} J_{\ell} \left( a^\dagger_\ell a_{\ell+1} +
a^\dagger_{\ell+1} a_\ell
 \right) +\sum_{\ell=0}^N h_{\ell} a^\dagger_\ell a_\ell.
\label{ham2}
\end{equation}
The Hamiltonian $H$ in (\ref{ham2}) describes a set of $N +1$
non--interacting (or free) fermions which hop between adjacent
sites of the lattice, and are subject to a non-uniform magenetic
field, denoted by $h_\ell$, $\ell=0, 1,\ldots, N$. Let
$|\ell\rangle$ denote a state in which there is a single fermion
at the site $\ell$ and all other sites are empty. Then the set of
states $\{|\ell\rangle\}$ forms a basis spanning the subspace
${\cal{S}}_1$. In this single--particle basis, the Hamiltonian $H$
is represented by the matrix
\begin{equation}
\left(%
\begin{array}{ccccc}
  h_0      & J_0   & 0          & \cdots  & 0       \\
  J_0    & h_1      & J_1        & \cdots  & 0       \\
  0      & J_1    & h_2          & \cdots  & 0       \\
  \vdots & \vdots & \vdots    & \ddots  & J_{N-1} \\
  0      & 0      & 0            & J_{N-1} & h_N       \\
\end{array}%
\right). \label{eq_singham}
\end{equation}
The dynamics of the register is completely determined by the
eigenvalues and eigenvectors of the above matrix. Let us denote
the energy eigenvalues of the matrix by $E_{k}$, where $k =
0,1,\ldots, N$, and the corresponding energy eigenfunction by
$\phi_{k}(\ell)$ (where $\ell \in \{0,1,\ldots, N\}$). The latter
corresponds to a single fermion at the site $\ell$ of the chain.
In the $M$--fermion sector, the energy of $M$ fermions occupying
orbitals $ 0 \leq k_1 < \cdots < k_M \leq N $ is then given by
\begin{equation}
    E_{k_1, \ldots,k_M} = \sum_{i=1}^M E_{k_i}
\end{equation}
and the corresponding $M$-particle energy eigenfunction can be
written as the Slater determinant
\begin{equation}
\Phi_{k_1,\ldots ,k_M}(\ell_1, \ldots, \ell_M) =
\frac{1}{\sqrt{M!}} \left|
\begin{matrix}
\phi_{k_1}(\ell_1) & \cdots & \phi_{k_1}(\ell_M)\\
 \vdots&  \ddots &  \vdots &\\
\phi_{k_M}(\ell_1) & \cdots & \phi_{k_M}(\ell_M)
\end{matrix}
\hspace{-0.2cm}\right| \label{eq_wavefunc}
\end{equation}
The eigenfunction $\Phi_{k_1,\ldots ,k_M}(\ell_1,\ldots, \ell_M)$
is completely antisymmetric. Let us now see how this eigenfunction
is related to the wavefunction of the quantum register.

In the subspace ${\cal{S}}_M$, the wavefunction of the register
$\Psi(s_0, \ldots, s_N)$ can be expressed as $\Psi(\ell_1,\ldots,
\ell_M)$, where $\ell_1,\ldots, \ell_M$ label the qubits which
have bit values equal to $1$. Each of the remaining qubits have
bit value $0$. In other words, the value of $\Psi(\ell_1,\ldots,
\ell_M)$ gives the probability amplitude that the qubits located
at the sites $\ell_1, \ell_2, \ldots, \ell_M$ represent binary `1'
and all other qubits represent binary `0'. Note that the
wavefunction $\Psi(\ell_1, \ldots, \ell_M)$ is symmetric under an
interchange of its labels, and is hence bosonic. It can, however,
be expressed in terms of the fermionic wavefunctions $\Phi_{k_1,
\ldots ,k_M}(\ell_1, \ldots, \ell_M)$ in the following manner: In
the sector $\ell_1 <\ell_2 <\ldots < \ell_M$, the wavefunction of
the register corresponding to the energy eigenvalue
$E_{k_1,\ldots, k_M}$ is set equal to the fermionic eigenfunction
$\Phi_{k_1,\ldots, k_M}(\ell_1, \ldots, \ell_M)$:
\begin{eqnarray}
\Psi(\ell_1,\ldots, \ell_M) &\equiv& \Psi_{k_1,\ldots, k_M}(\ell_1,\ldots, \ell_M)\\
&=& \Phi_{k_1,\ldots, k_M}(\ell_1,\ldots, \ell_M).
\end{eqnarray}
In the other sectors the two differ by the sign giving the parity
of the permutation required to reshuffle the arguments in
increasing order.

A Hamiltonian is said to be {\it mirror periodic} if it satisfies
\begin{equation}
e^{-i T{H}} \Psi(\ell_1,\ldots,\ell_M) =(\pm 1) \Psi (N-\ell_1,
\ldots ,N-\ell_M), \label{relation}
\end{equation}
for each $1 \leq M \leq N$, the sign depending on $M$ and $N$
only. Since $\Psi(\ell_1, \ldots, \ell_M)$ is bosonic, we can
choose $\ell_1 <\ell_2 <\ldots < \ell_M$. Now,

\begin{equation}
 e^{-i T{H}} \Psi(\ell_1,\ldots,\ell_M) =e^{-i T{H}} \Phi_{k_1,
\ldots, k_M}(\ell_1,\ldots,\ell_M). \label{relation-2}
\end{equation}

Our aim is to find Hamiltonians $H$ for which the RHS of
eq.~(\ref{relation-2}) is given by $\Phi_{k_1, \ldots,
k_M}(N-\ell_1, \ldots, N-\ell_M)$. This would imply mirror
periodicity of $H$ since

\begin{equation}
\begin{split}
\Phi_{k_1, \ldots, k_M}&(N-\ell_1, \ldots, N-\ell_M)=\\
&(-1)^{\frac{M(M-1)}{2}}\Psi(N-\ell_1, \ldots, N-\ell_M),
\end{split}
\end{equation}
by the above discussion.

The mirror periodicity with period $T$ implies periodicity with
period $2T$, which in turn implies that for all $k$ the quantity
$2 T E_k $ is an integer multiple of $2\pi$ in units for which
$\hbar = 1$ and $\phi_k(N-\ell) = \pm\phi_k(\ell)$.

We found two families of mirror periodic Hamiltonians: one
{\bf{(A)}} with linear spectrum and the other {\bf{(B)}} with
quadratic spectrum. An alternative proof of mirror periodicity for
the case {\bf{(A)}}, in the single-particle sector, was given by
Christandl {\it et al.}~\cite{CDEL04}. The proof relied on
identifying the Hamiltonian operator with the generator of space
rotations and employed group theoretical methods. In this paper we
recognize that the mirror periodicity extends to all
multi-particle sectors and that it is also shared by another
finite quantum chain with eigenfunctions given by Hahn
polynomials. Let us now discuss cases {\bf{(A)}} and {\bf{(B)}} in
detail:

{\bf (A)} The quantum chain with linear spectrum $P(k) = k$ has
eigenfunctions $\phi_k(\ell)$ proportional to Krawtchouk
polynomials. This polynomial basis has been used by Atakishiev
{\it et al.}~\cite{AJNW} to construct finite quantum chains
admitting periodic solutions.

The Krawtchouk quantum chain which is mirror periodic
of period $T = \pi$ has couplings
\begin{equation}
J_{\ell} = \sqrt{(\ell+1)(N-\ell)} \quad ; \quad h_\ell = 0,
\end{equation}
The Krawtchouk polynomials are defined in terms of the hypergeometric
functions $F$ as
\begin{equation}
K_k (\ell,p, N) = \; {_2F_1} \left(
\begin{matrix}
-k, -\ell \\
-N
\end{matrix}
\bigg\lvert \frac{1}{p}1 \right)
\end{equation}
where $k = 0, 1, 2, ..., N$. The energy eigenfunctions are
\begin{equation}
\phi_k(\ell) = c_k \sqrt{w(\ell)} K_k\left(\ell, {1\over 2}, N
\right)
\end{equation}
where $c_k$ and $w(\ell)$ are given by
\begin{equation} c_k = \sqrt{ (-N)_k \over (-1)^k
k! }, \qquad w(\ell) = \frac{1}{2^N} {N \choose \ell}.
\label{eq_wt1}
\end{equation}
The corresponding eigenvalues are $ E_k = - k.$
In the definitions above we have used the Pochhammer symbol,
$(N)_k$ defined as
\begin{equation}
(N)_k = N(N+1) \ldots (N+k-1), \;\;k=1,2,3,\ldots
\end{equation}
with $(N)_0 = 1$, and the generalized binomial symbol expressed in
terms of the $\Gamma$ function as
\begin{equation}
\left(
\begin{matrix}
N \\
\ell
\end{matrix}
\right) \;=\; {\Gamma(N+1) \over \Gamma(N-\ell+1) \Gamma(\ell+1)}.
\end{equation}
For a more comprehensive description of the Krawtchouk polynomials we
refer to \cite{Al-Salam}.

The energy eigenfunctions satisfy the property of reflection symmetry
(or antisymmetry):
\begin{equation}
\phi_k(N-\ell) = (-1)^k \phi_k(\ell) \label{eq_symm}
\end{equation}
for all $\ell$ and all $k=0,1,2,...,N$. This follows from the
following property of the Krawtchouk polynomials:
\begin{equation}
K_k \left( N- \ell; \frac{1}{2}, N \right) = (-1)^k K_k
\left(\ell; \frac{1}{2}, N \right),
\end{equation}
and the fact that the weight function in (\ref{eq_wt1}) is
symmetric. The phases $(-1)^k$ in (\ref{eq_symm}) perfectly offset
the dynamical phases acquired after a time period $T =\pi$, as
$\exp\left(- i T E_k) \right) = (-1)^k$. This shows that the chain
defined by the Hamiltonian corresponding to the Krawtchouk
polynomials is mirror symmetric with period $\pi$.

The dynamics of the Krawtchouk quantum chain of $N+1$ sites is the
same as that of a spin $s = N/2$ particle governed by the
Hamiltonian $H_s = 2 {\bf s}_x$. This Hamiltonian acts as follows
on the basis vectors $\lvert m \rangle, m = -s,...,s$:
\begin{equation}
H_{s} \lvert m \rangle =
R(m) \lvert m - 1 \rangle \nonumber +
L(m) \lvert m + 1 \rangle,
\end{equation}
where
\bea
R(m) \hskip-0.2cm &=& \hskip-0.2cm \sqrt{s(s+1) - m(m-1)}
\nonumber   \\
L(m) \hskip-0.2cm &=& \hskip-0.2cm \sqrt{s(s+1) - m(m+1)}.
\nonumber \eea It is possible to establish a relation between
$H_s$ and a mirror-periodic Krawtchouk chain of $N+1 = 2 s+1$
sites. This is done by identifying the state $\lvert \ell\rangle$
corresponding to a single particle occupying the site $\ell$ with
the state $\lvert m\rangle$, where $m = s - \ell$. In this case,
$R(m)$ reduces to $J_\ell$ and $L(m)$ reduces to $J_{\ell-1}$,
showing that the spin Hamiltonian $H_s$ is equivalent to the
mirror periodic Krawtchouk Hamiltonian.

{\bf (B)} One can use Hahn polynomials to find a family of
mirror periodic quantum chains whose period is an integer multiple
of $\pi$ with quadratic spectrum
$E_k = k(k + 2 \alpha + 1)$, where $\alpha$ is of the form
\begin{equation}
\alpha = {2 p + 1 \over 2 q}
\label{eq_alpha}
\end{equation}
where $p, q$ are integers with $q \neq 0$. The couplings are
\begin{equation}
J_{\ell} = \sqrt{ (\ell+1)( N - \ell ) \left( \alpha + N - \ell
\right) \left( \alpha + \ell + 1 \right) }
\end{equation}
and the Zeeman terms are given by
\begin{equation}
h_\ell = {N^2\over2} + (\alpha+1) N - 2 \left( \ell - {N\over2}
\right)^2.
\end{equation}

This model has eigenfunctions $\phi_k(\ell)$ given by Hahn
polynomials. The Hahn polynomials are defined in terms of the
hypergeometric functions $F$ as
\begin{equation}
Q_k(\ell; \alpha, \beta, N) = \; {_3F_2} \left(
\begin{matrix}
-k, k+\alpha+\beta+1, \ell \\
\alpha+1, -N
\end{matrix}
\bigg\lvert 1 \right)
\end{equation}
where $k = 0, 1, 2, ..., N$. The energy eigenfunctions of the
Hamiltonian (\ref{eq_spinham}) are given by
\begin{equation}
\phi_k(\ell) = c_k \sqrt{w(\ell)} Q_k\left(\ell; \alpha, \alpha, N
\right)
\end{equation}
where $c_k$ is the constant
\begin{equation}
c_k = \sqrt{ (2k + 2\alpha+1) (N!)^2 \over
\left( k + 2\alpha + 1\right)_{N+1} k! (N-k)!},
\label{const}
\end{equation}
and $w(\ell)$ is the weight function
\begin{equation}
w(\ell) = \left(
\begin{matrix}
\alpha + \ell \\
\ell
\end{matrix}
\right) \left(
\begin{matrix}
\alpha + N - \ell \\
N - \ell
\end{matrix}
\right). \label{eq_weight}
\end{equation}
For further details on Hahn polynomials
see \cite{Al-Salam}.

To show that the $\phi_k(\ell)$ are either reflection symmetric or
anti-symmetric, we notice that
\begin{equation}
Q_k(N- \ell; \alpha, \alpha, N) = (-1)^k Q_k(\ell; \alpha, \alpha,
N),
\end{equation}
and that the weight function in (\ref{eq_weight}) is symmetric.
Hence, $ \phi_k(N-\ell) = (-1)^k \phi_k(\ell), $ for all $\ell$
and all $k=0,1,2,...,N$. If $\alpha$ satisfies (\ref{eq_alpha})
and $T = q\pi$, the phases $(-1)^k$ perfectly offset the dynamical
phases. In fact
\begin{equation}
\exp ( - i T E_k ) = \exp ( - i \pi [q (k^2 + k) + (2 p + 1)k] ) =
(-1)^k
\end{equation}
since $k^2 + k$ is even for all $k = 0, ..., N$.
This shows that the Hahn chain is mirror periodic of
period $T = q\pi$.

The Hahn chain Hamiltonian in the special case $q=1$, i.e.
when $\alpha = (2p+1) / 2$ is half-integer, is related
to atomic Hamiltonians with
${\bf L}\cdot {\bf S}$ coupling.
Consider the Hamiltonian
\begin{equation}
H_{LS} = {\bf L}\cdot {\bf S}
\label{eq_hls}
\end{equation}
restricted to the sector with fixed total angular momentum $L$,
total spin $S$ and with projections along a fixed axis adding up
to zero, i.e. $M = M_L + M_S = 0$. The Hamiltonian in
(\ref{eq_hls}) acts as follows on the basis vectors $\lvert M_S
\rangle \equiv \lvert L, S; M_L, M_S \rangle$:
\begin{equation}
H_{LS} \lvert M_S \rangle = D \lvert M_S \rangle + R \lvert M_S -
1 \rangle  + L \lvert M_S + 1 \rangle
\end{equation}
where
\begin{eqnarray}
D \hspace{-0.1cm}&\equiv& D(M_S) = - M_S^2 \\
R  \hspace{-0.1cm}&\equiv& R(M_S) = {1\over 2}\sqrt{(L + M_S)(L - M_S + 1)} \nonumber\\
&&\times \sqrt{(S + M_S) (S - M_S + 1) } \\
L  \hspace{-0.1cm}&\equiv& L(M_S) = {1\over 2}\sqrt{(L - M_S)(L +
M_S + 1)}\nonumber\\
&&\times \sqrt{(S - M_S) (S + M_S + 1) }.
\end{eqnarray}

Assuming that $S < L$ and that $S$ is a half-integer, it is
possible to establish a relation between $H_{LS}$ and a
mirror-periodic Hahn chain of $N = 2 S$ sites and $\alpha = L -
S$. This is done by identifying the state $\lvert \ell\rangle$
corresponding to a single particle occupying the site $\ell$ with
the state $\lvert M_S\rangle$, where $M_S = S - \ell$. We find
$R(M_S) = {1\over 2} J_\ell$, $L(M_S) = {1\over 2} J_{\ell-1}$ and
$D(M_S) = {1\over 2} h_\ell + {\rm const}$. This shows that the LS
coupling Hamiltonian is proportional to a mirror periodic Hahn
Hamiltonian up to a constant energy shift.

In conclusion, in this Letter we have demonstrated how to simplify
transfer of data in quantum registers by implementing a mirror
inversion of a quantum state with respect to the centre of the
register. Our construction is especially appealing as it requires
no dynamical control over individual qubits but only
pre-engineered inter-qubit couplings. If, however, individual
control of the interactions is available then the mirror inversion
operation can be performed on any substring of qubits in the
register. In this case a sequence of mirror inversions can
generate any permutation of a quantum state of the involved
qubits.

\begin{acknowledgments}
This work was supported in part by a grant from the Cambridge-MIT
Institute, A$^*$Star Grant No.\ 012-104-0040 and the EU under
project RESQ (IST-2001-37559). CA was supported by the National
Science and Engineering Council of Canada under grant
RGPIN-171149. MC acknowledges the support of a DAAD
Doktorandenstipendium. We would like to thank H. Haselgrove for
pointing out an error in the previous version.
\end{acknowledgments}


\begin{thebibliography}{16}
\expandafter\ifx\csname
natexlab\endcsname\relax\def\natexlab#1{#1}\fi
\expandafter\ifx\csname bibnamefont\endcsname\relax
  \def\bibnamefont#1{#1}\fi
\expandafter\ifx\csname bibfnamefont\endcsname\relax
  \def\bibfnamefont#1{#1}\fi
\expandafter\ifx\csname citenamefont\endcsname\relax
  \def\citenamefont#1{#1}\fi
\expandafter\ifx\csname url\endcsname\relax
  \def\url#1{\texttt{#1}}\fi
\expandafter\ifx\csname
urlprefix\endcsname\relax\def\urlprefix{URL }\fi
\providecommand{\bibinfo}[2]{#2}
\providecommand{\eprint}[2][]{\url{#2}}

\bibitem[{\citenamefont{Deutsch}(1989)}]{Deu89}
\bibinfo{author}{\bibfnamefont{D.}~\bibnamefont{Deutsch}},
  \bibinfo{journal}{Proc. R. Soc. of Lond. A} \textbf{\bibinfo{volume}{425}},
  \bibinfo{pages}{73} (\bibinfo{year}{1989}).

\bibitem[{\citenamefont{Mandel et~al.}(2003)\citenamefont{Mandel, Greiner,
  Widera, Rom, H\"ansch, and Bloch}}]{MGWRHB03}
\bibinfo{author}{\bibfnamefont{O.}~\bibnamefont{Mandel}},
  \bibinfo{author}{\bibfnamefont{M.}~\bibnamefont{Greiner}},
  \bibinfo{author}{\bibfnamefont{A.}~\bibnamefont{Widera}},
  \bibinfo{author}{\bibfnamefont{T.}~\bibnamefont{Rom}},
  \bibinfo{author}{\bibfnamefont{T.~W.} \bibnamefont{H\"ansch}},
  \bibnamefont{and} \bibinfo{author}{\bibfnamefont{I.}~\bibnamefont{Bloch}},
  \bibinfo{journal}{Nature} \textbf{\bibinfo{volume}{425}},
  \bibinfo{pages}{937} (\bibinfo{year}{2003}).

\bibitem[{\citenamefont{Ekert}(1995)}]{Eke95e}
\bibinfo{author}{\bibfnamefont{A.}~\bibnamefont{Ekert}}, in
  \emph{\bibinfo{booktitle}{Advances in Quantum Phenomena}}, edited by
  \bibinfo{editor}{\bibnamefont{E.G.Beltrametti}} \bibnamefont{and}
  \bibinfo{editor}{\bibnamefont{J-M.Levy-Leblond}} (\bibinfo{publisher}{Plenum
  Press}, \bibinfo{year}{1995}), vol. \bibinfo{volume}{347} of
  \emph{\bibinfo{series}{NATO ASI Series B: Physics}}, pp.
  \bibinfo{pages}{243--262}.

\bibitem[{\citenamefont{Barenco et~al.}(1995)\citenamefont{Barenco, Deutsch,
  Ekert, and Jozsa}}]{BDEJ95}
\bibinfo{author}{\bibfnamefont{A.}~\bibnamefont{Barenco}},
  \bibinfo{author}{\bibfnamefont{D.}~\bibnamefont{Deutsch}},
  \bibinfo{author}{\bibfnamefont{A.}~\bibnamefont{Ekert}}, \bibnamefont{and}
  \bibinfo{author}{\bibfnamefont{R.}~\bibnamefont{Jozsa}},
  \bibinfo{journal}{Phys. Rev. Lett.} \textbf{\bibinfo{volume}{74}},
  \bibinfo{pages}{4083} (\bibinfo{year}{1995}).

\bibitem[{\citenamefont{Kane}(1998)}]{Kan98a}
\bibinfo{author}{\bibfnamefont{B.~E.} \bibnamefont{Kane}},
  \bibinfo{journal}{Nature} \textbf{\bibinfo{volume}{393}},
  \bibinfo{pages}{133} (\bibinfo{year}{1998}).

\bibitem[{\citenamefont{Loss and DiVincenzo}(1998)}]{LD98a}
\bibinfo{author}{\bibfnamefont{D.}~\bibnamefont{Loss}} \bibnamefont{and}
  \bibinfo{author}{\bibfnamefont{D.~P.} \bibnamefont{DiVincenzo}},
  \bibinfo{journal}{Phys. Rev. A} \textbf{\bibinfo{volume}{57}},
  \bibinfo{pages}{120} (\bibinfo{year}{1998}).

\bibitem[{\citenamefont{Cory et~al.}(1996)\citenamefont{Cory, Fahmy, and
  Havel}}]{CFH96a}
\bibinfo{author}{\bibfnamefont{D.~G.} \bibnamefont{Cory}},
  \bibinfo{author}{\bibfnamefont{A.~F.} \bibnamefont{Fahmy}}, \bibnamefont{and}
  \bibinfo{author}{\bibfnamefont{T.~F.} \bibnamefont{Havel}}, in
  \emph{\bibinfo{booktitle}{Proceedings of the 4th Workshop on Physics and
  Computation}}, edited by
  \bibinfo{editor}{\bibfnamefont{T.}~\bibnamefont{Toffoli}}
  \bibnamefont{et~al.} (\bibinfo{publisher}{New England Complex Systems
  Institute}, \bibinfo{address}{Boston, Massachusetts}, \bibinfo{year}{1996}),
  pp. \bibinfo{pages}{87--91}.

\bibitem[{\citenamefont{Gershenfeld and Chuang}(1997)}]{GC97a}
\bibinfo{author}{\bibfnamefont{N.}~\bibnamefont{Gershenfeld}} \bibnamefont{and}
  \bibinfo{author}{\bibfnamefont{I.~L.} \bibnamefont{Chuang}},
  \bibinfo{journal}{Science} \textbf{\bibinfo{volume}{275}},
  \bibinfo{pages}{350} (\bibinfo{year}{1997}).

\bibitem[{\citenamefont{DiVincenzo et~al.}(2000)\citenamefont{DiVincenzo,
  Bacon, Kempe, Burkard, and Whaley}}]{DBKB00}
\bibinfo{author}{\bibfnamefont{D.~P.} \bibnamefont{DiVincenzo}},
  \bibinfo{author}{\bibfnamefont{D.}~\bibnamefont{Bacon}},
  \bibinfo{author}{\bibfnamefont{J.}~\bibnamefont{Kempe}},
  \bibinfo{author}{\bibfnamefont{G.}~\bibnamefont{Burkard}}, \bibnamefont{and}
  \bibinfo{author}{\bibfnamefont{K.~B.} \bibnamefont{Whaley}},
  \bibinfo{journal}{Nature} \textbf{\bibinfo{volume}{408}},
  \bibinfo{pages}{339} (\bibinfo{year}{2000}).

\bibitem[{\citenamefont{Benjamin}(2002)}]{Ben02}
\bibinfo{author}{\bibfnamefont{S.~C.} \bibnamefont{Benjamin}},
  \bibinfo{journal}{Phys. Rev. Lett.} \textbf{\bibinfo{volume}{88}},
  \bibinfo{pages}{017904} (\bibinfo{year}{2002}).

\bibitem[{\citenamefont{Zhou et~al.}(2002)\citenamefont{Zhou, Zhou, Guo, and
  Feldman}}]{ZZGF02}
\bibinfo{author}{\bibfnamefont{X.}~\bibnamefont{Zhou}},
  \bibinfo{author}{\bibfnamefont{Z.-W.} \bibnamefont{Zhou}},
  \bibinfo{author}{\bibfnamefont{G.-C.} \bibnamefont{Guo}}, \bibnamefont{and}
  \bibinfo{author}{\bibfnamefont{M.}~\bibnamefont{Feldman}},
  \bibinfo{journal}{Phys. Rev. Lett.} \textbf{\bibinfo{volume}{89}},
  \bibinfo{pages}{197903} (\bibinfo{year}{2002}).

\bibitem[{\citenamefont{Benjamin and Bose}(2003)}]{BB03}
\bibinfo{author}{\bibfnamefont{S.~C.} \bibnamefont{Benjamin}} \bibnamefont{and}
  \bibinfo{author}{\bibfnamefont{S.}~\bibnamefont{Bose}},
  \bibinfo{journal}{Phys. Rev. Lett.} \textbf{\bibinfo{volume}{90}},
  \bibinfo{pages}{247901} (\bibinfo{year}{2003}).

\bibitem[{\citenamefont{Lieb and Wu}(1968)}]{LiebWu}
\bibinfo{author}{\bibfnamefont{E.}~\bibnamefont{Lieb}} \bibnamefont{and}
  \bibinfo{author}{\bibfnamefont{F.}~\bibnamefont{Wu}}, \bibinfo{journal}{Phys.
  Rev. Lett.} \textbf{\bibinfo{volume}{20}},
  \bibinfo{pages}{1445}  (\bibinfo{year}{1968}).

\bibitem[{\citenamefont{Christandl et~al.}(2004)\citenamefont{Christandl,
  Datta, Ekert, and Landahl}}]{CDEL04}
\bibinfo{author}{\bibfnamefont{M.}~\bibnamefont{Christandl}},
  \bibinfo{author}{\bibfnamefont{N.}~\bibnamefont{Datta}},
  \bibinfo{author}{\bibfnamefont{A.}~\bibnamefont{Ekert}}, \bibnamefont{and}
  \bibinfo{author}{\bibfnamefont{A.~J.} \bibnamefont{Landahl}},
  \bibinfo{journal}{Phys. Rev. Lett.} \textbf{\bibinfo{volume}{92}},
  \bibinfo{pages}{187902} (\bibinfo{year}{2004}).

\bibitem[{\citenamefont{Atakishiev et~al.}(1998)\citenamefont{Atakishiev,
  Jafarov, Nagiyev, and Wolf}}]{AJNW}
\bibinfo{author}{\bibfnamefont{A.~N.} \bibnamefont{Atakishiev}},
  \bibinfo{author}{\bibfnamefont{E.}~\bibnamefont{Jafarov}},
  \bibinfo{author}{\bibfnamefont{S.}~\bibnamefont{Nagiyev}}, \bibnamefont{and}
  \bibinfo{author}{\bibfnamefont{K.}~\bibnamefont{Wolf}},
  \bibinfo{journal}{Rev.Mex.Fis.}  \textbf{\bibinfo{volume}{44}},
  \bibinfo{pages}{235} (\bibinfo{year}{1998}).

\bibitem[{\citenamefont{Al-Salam}(1990)}]{Al-Salam}
\bibinfo{author}{\bibfnamefont{W.}~\bibnamefont{Al-Salam}},
  \bibinfo{journal}{In : {O}rthogonal {P}olynomials : {T}heory and {P}ractice
  (ed. {P}. {N}evai), {K}luwer {A}cademic {P}ublishers, {D}ordrecht}
  (\bibinfo{year}{1990}).

\end{thebibliography}

\end{document}